\documentclass[notitlepage,pre,proof,longbibliography]{revtex4-1}%
\usepackage{amssymb}
\usepackage{amsfonts}
\usepackage{amsmath}
\usepackage{xcolor}
\usepackage{ulem}
\usepackage{graphicx}%
\providecommand{\U}[1]{\protect\rule{.1in}{.1in}}
\begin{document}
\title{Explicitly stable Fundamental Measure Theory models for classical density functional theory.}
\author{James F. Lutsko}
\homepage{http://www.lutsko.com}
\email{jlutsko@ulb.ac.be}
\affiliation{Center for Nonlinear Phenomena and Complex Systems CP 231, Universit\'{e}
Libre de Bruxelles, Blvd. du Triomphe, 1050 Brussels, Belgium}

\pacs{PACS number}

\begin{abstract}
The derivation of the state of the art tensorial versions of Fundamental
Measure Theory (a form of classical Density Functional Theory for hard spheres) are re-examined in the light of the recently introduced concept
of global stability of the density functional based on its boundedness (Lutsko
and Lam, Phys. Rev. E 98, 012604 (2018)). It is shown that within the present
paradigm, explicitly stability of the functional can be achieved only at the
cost of giving up accuracy at low densities. It is argued that this is an
acceptable trade-off since the main value of DFT lies in the study of dense
systems. Explicit calculations for a wide variety of systems shows that a
proposed explicitly stable functional is competitive in all ways with the
popular White Bear model while sharing some of its weaknesses  when
applied to non-close-packed solids.

\end{abstract}
\maketitle

\section{Introduction}
Classical density functional theory (cDFT) has become an important tool for
studying nanoscale phenomena such as the solvation energy of large
ions\cite{Borgis}, crystallization\cite{Lutsko_HCF}, confinement-induced
polymorphism\cite{LamLutsko} and wetting\cite{Evans23901}, to name a few. Key
to this utility is the ability of cDFT to accurately describe molecular-scale
correlations ultimately arising from excluded volume effects. To do this,
state of the art cDFT models incorporate as a fundamental building block
highly sophisticated functionals for hard-spheres that have been developed
over the past 30 years. Indeed, the determination of the \textit{exact}
functional for hard spheres in one dimension (hard-rods)\ by
Percus\cite{Percus1, Percus2, Percus3} in the 1970's was a key milestone in
the development of cDFT and served to inspire the modern class of models known
as Fundamental Measure Theory (FMT)\cite{Rosenfeld1}. Because of the central role that these
play in important applications, the perfection of these models, in so far as
possible, remains an important subject of research.

All forms of finite temperature DFT\ (quantum and classical) are based on
fundamental theorems asserting the existence of a functional $\Lambda\left[
\rho;\phi\right]  $ of the local density $\rho\left(  \mathbf{r}\right)  $ and
any one-body fields $\phi\left(  \mathbf{r}\right)  $ having the property that
it is uniquely minimized by the equilibrium density distribution for the
system, $\rho_{\text{eq}}\left(  \mathbf{r}\right)  $, and that $\Lambda
\left[  \rho_{\text{eq}};\phi\right]  $ is the system's (grand-canonical) free
energy (for overviews of cDFT see \cite{Evans1979, lutsko:acp}). In general,
the dependence on the field is trivial and is separated off by writing
$\Lambda\left[  \rho;\phi\right]  =F\left[  \rho\right]  +\int\rho\left(
\mathbf{r}\right)  \phi\left(  \mathbf{r}\right)  d\mathbf{r}$ where the
first, field-independent, term $F$ is called the Helmholtz functional. This in
turn can be divided into the sum of a (known) ideal-gas term, $F_{\text{id}%
}\left[  \rho\right]  $ and an (unknown) excess term $F_{\text{ex}}\left[
\rho\right]  $ which is the main focus of attention. The FMT model for the excess term was originally
developed by Rosenfeld\cite{Rosenfeld1} based on ideas from Scaled Particle
Theory. Since then, other approaches to the subject have been explored and
one, so-called dimensional crossover\cite{RSLT, Tarazona1997a}, has proven
particularly fruitful. This starts with a general ansatz for $F_{\text{ex}%
}\left[  \rho\right]  $ and then attempts to work out its details by demanding
that the theory in , e.g., three dimensions reproduce exact lower-dimensional
functionals when suitably restricted. This turns out to lead in a
straightforward manner to FMT and was the inspiration for the most
sophisticated modern "White Bear" functionals. These give a very good
description of inhomogeneous hard-sphere systems including the freezing transition.

While the White Bear functionals\cite{WBI,WBII} are the best overall models
available for three dimensional systems, they are not without flaws. For
example, it has been known for some time that they do not give a very
satisfactory description of non-close packed crystal
structures\cite{Lutsko_FCC}. More importantly, it has recently been
demonstrated\cite{LutskoLam} that they are unstable in the sense that the free
energy is unbounded from below leading to the possibility that no global
minimum exists. This means that successful applications of these models
involve local (meta stable) minima which, while giving physically reasonable
results, is in violation of the fundamental theorems of cDFT which say that the
equilibrium state must be a global minimum of the free energy functional. The
issue is not just philosophical: the local minima can be explored in, e.g.,
perfect crystals due to the high symmetry but in more general applications
such as those in Refs.\cite{Lutsko_HCF} and \cite{LamLutsko} cited above, a
complete lack of symmetry leaves the calculations vulnerable to such
unphysical global minima. Another conceptual drawback is that to achieve a
good quantitative of dense liquids, the functionals that come out of the
dimensional-crossover path are heuristically modified in a way that
invalidates the recovery of the low-dimensional results that were their
motivation in the first place.

In this paper, I revisit the concept of dimensional crossover while paying
particular attention to the question of stability of the functionals. In the
next Section, it is shown that within the current paradigm of FMT, including
the usual but not necessary limits on the complexity of the functionals,
explicit global stability can only be achieved by giving up some accuracy at
low densities. This seems a fair trade-off as the main utility of cDFT lies in
modeling correlations of dense systems for which the low-density accuracy is
not so important. The following section presents explicit calculations using
an explicitly stable model and shows that it is in most ways comparable in
practical accuracy to the most sophisticated White Bear models across a wide
range of systems: homogeneous fluid, inhomogeneous fluid near a wall, and
various solid structures. The last section summarized these results and
discusses possibilities for further development in this direction.

\section{Fundamental Measure Theory and Dimensional Crossover}

\subsection{The standard form of FMT}

In FMT the excess functional for a single hard-sphere species with radius $R$
and diameter $\sigma=2R$, is written as
\begin{equation}
F_{\text{ex}}\left[  \rho\right]  =\int\left\{  \Phi_{1}\left(
\overrightarrow{n}\left(  \mathbf{r};\left[  \rho\right]  \right)  \right)
+\Phi_{2}\left(  \overrightarrow{n}\left(  \mathbf{r};\left[  \rho\right]
\right)  \right)  +\Phi_{3}\left(  \overrightarrow{n}\left(  \mathbf{r}%
;\left[  \rho\right]  \right)  \right)  \right\} d\mathbf{r}
\end{equation}
where $\rho(\mathbf{r})$ is the local number density and the array of fundamental measures $\overrightarrow{n}$ include the local
packing fraction $\eta\left(  \mathbf{r};\left[  \rho\right]  \right)  $, and
the scalar, vector and tensor surface measure $s\left(  \mathbf{r};\left[
\rho\right]  \right)  $, $\mathbf{v}\left(  \mathbf{r};\left[  \rho\right]
\right)  $ and $\mathbf{T}\left(  \mathbf{r};\left[  \rho\right]  \right)  $
respectively:%
\begin{align}
\eta\left(  \mathbf{r};\left[  \rho\right]  \right)   &  =\int\Theta\left(
R-r_{1}\right)  \rho\left(  \mathbf{r-r}_{1}\right)  d\mathbf{r}_{1}\\
\left(
\begin{array}
[c]{c}%
s\left(  \mathbf{r};\left[  \rho\right]  \right) \\
\mathbf{v}\left(  \mathbf{r};\left[  \rho\right]  \right) \\
\mathbf{T}\left(  \mathbf{r};\left[  \rho\right]  \right)
\end{array}
\right)   &  =\int\left(
\begin{array}
[c]{c}%
1\\
\widehat{\mathbf{r}}_{1}\\
\widehat{\mathbf{r}}_{1}\widehat{\mathbf{r}}_{1}%
\end{array}
\right)  \delta\left(  R-r_{1}\right)  \rho\left(  \mathbf{r-r}_{1}\right)
d\mathbf{r}_{1}\nonumber
\end{align}
In the expression for $F_{\text{ex}}$, the first contribution $\Phi_{1}\left(
\overrightarrow{n}\right)  =-\frac{1}{S_{D}\sigma^{D}}s\ln\left(
1-\eta\right)  $ where $S_{D}$ is the surface area of a sphere with unit
diameter in $D$ dimensions. For $D=1$, this is precisely the exact result of
Percus and, in fact, FMT was developed as part of a general effort at the time
to extend this result to higher dimensions. Specializing to three dimensions,
the second term is $\Phi_{2}\left(  \overrightarrow{n}\right)  =\frac{1}%
{2\pi\sigma}\frac{s^{2}-v^{2}}{\left(  1-\eta\right)  }$ and the sum $\Phi
_{1}+\Phi_{2}$ has the remarkable property that when applied to a
one-dimensional density distribution, e.g. $\rho\left(  \mathbf{r}\right)
=\delta\left(  x\right)  \delta\left(  y\right)  \rho\left(  z\right)  $, it
reduces to the exact $D=1$ result. Note that such a density corresponds to a
3D system subject to an external field $\phi_{\varepsilon}\left(
\mathbf{r}\right)  $ which is infinite for $\left\vert x\right\vert
,\left\vert y\right\vert >\varepsilon R$ in the limit that $\varepsilon
\rightarrow0$:\ physically, this is a linear tube with radius slightly larger
than that of a hard-sphere.

\subsection{Global stability}

The overall stability of the functionals demands that they be bounded from
below: no density field can give rise to divergent negative $F_{\text{ex}%
}\left[  \rho\right]  $ as this would imply that the free energy of the
equilibrium system was divergently negative, which is unphysical. However, the
structure of FMT leaves this possibility open in that the functions $\Phi_{1}$
and $\Phi_{2}$ as well as all models for $\Phi_{3}$ have numerators that
depend on the surface densities divided by denominators that depend on the
local packing fraction (a volumetric term). If the numerators are negative at
some point $\mathbf{r}$\textbf{ }for some density field, e.g. if in $\Phi_{2}$
one had $s^{2}\left(  \mathbf{r}^{\ast};\left[  \rho\right]  \right)
-v^{2}\left(  \mathbf{r}^{\ast};\left[  \rho\right]  \right)  <0$ for some
$\rho\left(  \mathbf{r}\right)  $, then one could hold the density constant
constant on the sphere $\left(  \mathbf{r}-\mathbf{r}^{\ast}\right)
^{2}=R^{2}$ but increase it inside the sphere with the effect that the
numerator remains constant while the denominator, $1-\eta\left(
\mathbf{r}^{\ast};\left[  \rho\right]  \right)  $, will eventually become zero
giving rise to a divergence. The fact that the functional can diverge at one
point does not prove that it is unstable, but it allows for the possibility.
In fact, for both $\Phi_{1}$ and $\Phi_{2}$ this cannot occur: in the former
case we have only to note that $s\left(  \mathbf{r};\left[  \rho\right]
\right)  $ is positive semi-definite while the non-negativity in the latter
case will become apparent below. On the other hand, models for $\Phi_{3}$ have
long suffered from known instabilities\cite{RSLT} and suspected instabilities
(see the discussion in Ref. \cite{LutskoLam}). Thus, the object here will be
to re-examine the reasoning leading to the most useful current models for
$\Phi_{3}$ and to attempt to modify them so as to impose global stability.

\subsection{Dimensional Crossover}

\begin{figure}
[htp!]
\includegraphics[trim={0.5cm 8.5cm, 1.5cm 1.5cm}, clip=true,width=0.60\linewidth]{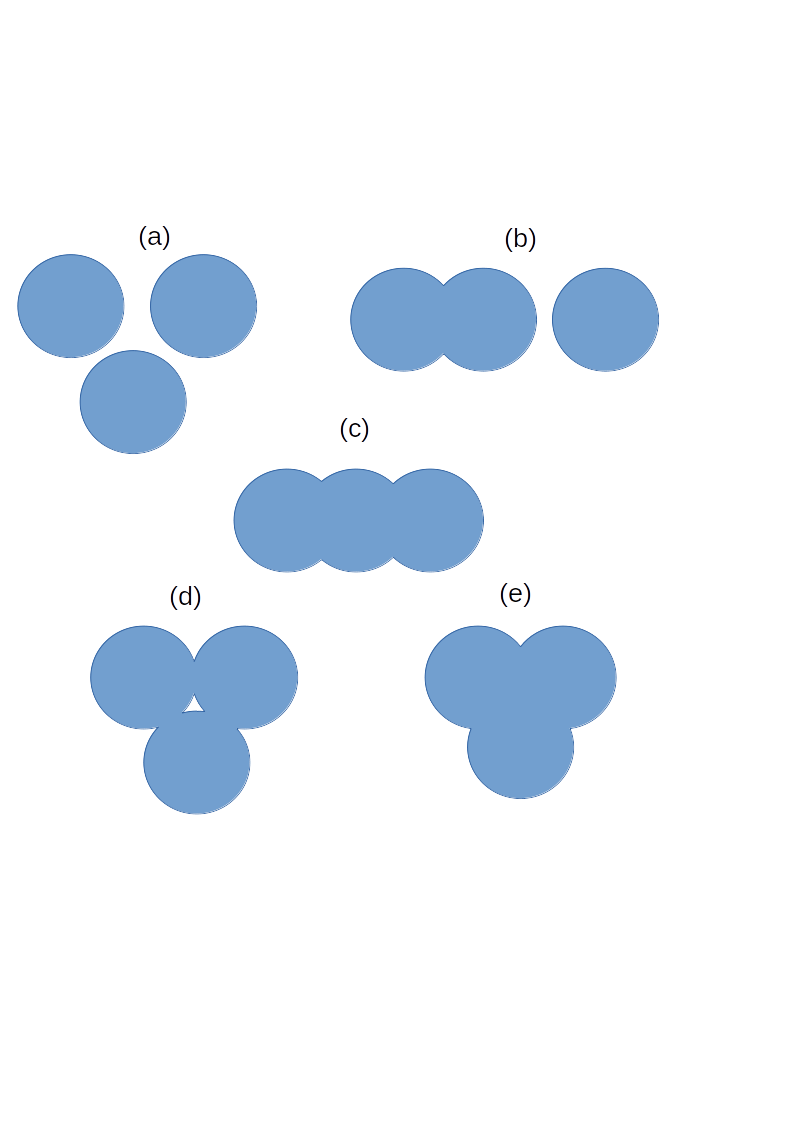}
\caption{The different classes of cavities made up of three spherical cavities. Each spherical cavity is just large enough to hold a single hard-sphere. The case of two cavities correspond to cases (a) and (b) with one of the detached cavities removed. Cases (c) and (d) cannot be captured by the FMT ansatz.}
\label{fig0}
\end{figure}

The idea of reproducing exact lower-dimensional results in a higher
dimensional system was termed by Rosenfeld and collaborators "dimensional
crossover"\cite{RSLT}. One of the only other cases for which exact results are
known, besides that of hard-rods in one dimension,  are the so-called 0D systems consisting of a cavity which is so small
that it can hold at most a single hard sphere (see Appendix \ref{app0D} for a derivation of the results used here). The minimal example would be a
spherical cavity with radius $R\left(  1+\varepsilon\right)  $ in the limit
that $\varepsilon\rightarrow0$ corresponding to a single point. In such a
case, the density can only take the form $\rho_{1}\left(  \mathbf{r}\right)
=N\delta\left(  \mathbf{r}\right)  $ with $0\leq N\leq1$ being the only
unknown (its value is determined by the chemical potential). The exact free
energy functional is easy to workout, $F_{\text{ex}}\left[  \rho\right]
=\Phi_{0}\left(  N[\rho]\right)  $ where $\Phi_{0}\left(  x\right)  =\left(
1-x\right)  \ln\left(  1-x\right)  -\left(  1-x\right)$ and $N[\rho] = \int \rho(\mathbf{r}) d\mathbf{r}$ is the average number of particles. Other closely
related examples are also accessible: for example, two such cavities that do
not overlap (a trivial generalization) or, more interestingly, two such cavities
which do overlap (see Fig.\ref{fig0}). In the latter case, the combined cavity
can still only hold a single hard sphere but the geometry allows a density
$\rho_{2}\left(  \mathbf{r}\right)  =N_{1}\delta\left(  \mathbf{r}%
-\mathbf{s}_{1}\right)  +N_{2}\delta\left(  \mathbf{r}-\mathbf{s}_{2}\right)
$ where $\mathbf{s}_{i}$ is the center of the $i$-th spherical cavity. In this
case, the unknown amplitudes must satisfy $0\leq N_{1}+N_{2}\leq1$ and the two
could be different if, e.g., the external field were different at the two
centers. If the two cavities overlap, then one finds that $F_{\text{ex}%
}\left[  \rho_{2}\right]  =\Phi_{0}\left(  N_{1}+N_{2}\right)  $ whereas if
they do not then $F_{\text{ex}}\left[  \rho_{2}\right]  =\Phi_{0}\left(
N_{1}\right)  +\Phi_{0}\left(  N_{2}\right)  $. In the case of three cavities,
the results are $F_{\text{ex}}\left[  \rho_{3}\right]  =\Phi_{0}\left(
N_{1}+N_{2}+N_{3}\right)  $, if the there is a non-zero mutual overlap of all
three, it is $F_{\text{ex}}\left[  \rho_{3}\right]  =\Phi_{0}\left(
N_{1}+N_{2}\right)  +\Phi_{0}\left(  N_{3}\right)  $ if the first two overlap
and neither intersects the third volume, and it is $F_{\text{ex}}\left[
\rho_{3}\right]  =\Phi_{0}\left(  N_{1}\right)  +\Phi_{0}\left(  N_{2}\right)
+\Phi_{0}\left(  N_{3}\right)  $ if none of them overlap. The results are more
complex for the case that cavities 1 and 3 both intersect cavity 2 but do not
intersect one another (e.g. a linear chain). This pattern holds for higher
numbers of cavities.

To use this information to recover FMT, Rosenfeld and
Tarazona\cite{Tarazona1997a} proposed that the excess functional be written as
a sum of terms,
\begin{equation}
F_{\text{ex}}\left[  \rho\right]  =F_{\text{ex}}^{\left(  1\right)  }\left[
\rho\right]  +F_{\text{ex}}^{\left(  2\right)  }\left[  \rho\right]  +...
\end{equation}
which are generated by the ansatz%
\begin{align}
F_{\text{ex}}^{\left(  1\right)  }\left[  \rho\right]   &  =\int
d\mathbf{r}\;\psi_{1}\left(  \eta\left(  \mathbf{r};\left[  \rho\right]
\right)  \right)  \int d\mathbf{r}_{1}\;\rho\left(  \mathbf{r-r}_{1}\right)
\delta\left(  R-r_{1}\right)  \delta\left(  R-r_{2}\right)  K_{1}\left(
\widehat{\mathbf{r}}_{1}\right)  ,\\
F_{\text{ex}}^{\left(  2\right)  }\left[  \rho\right]   &  =\int
d\mathbf{r}\;\psi_{2}\left(  \eta\left(  \mathbf{r};\left[  \rho\right]
\right)  \right)  \int d\mathbf{r}_{1}d\mathbf{r}_{2}\;\rho\left(
\mathbf{r-r}_{1}\right)  \rho\left(  \mathbf{r-r}_{2}\right)  \delta\left(
R-r_{1}\right)  \delta\left(  R-r_{2}\right)  K_{2}\left(  \widehat
{\mathbf{r}}_{1},\widehat{\mathbf{r}}_{2}\right)  ,\nonumber
\end{align}
and so forth. They then evaluated these for various 0D systems: a single
cavity, two cavities, etc. It is easy to see that even for the single cavity,
the second and all higher contributions will diverge unless the kernels,
$K_{n}\left(  \widehat{\mathbf{r}}_{1},...\widehat{\mathbf{r}}_{n}\right)  $
vanish whenever two of the arguments are equal: this is a fundamental
stability constraint on the construction of these models since, otherwise, the
densities will yield the squares and higher powers of Dirac delta-functions
leading to undefined results (see Appendix \ref{appB}). Since the kernels in the ansatz are scalars,
they must be constructed from the scalar products of its arguments so it is convenient to write them as $K_{1}\left(  \widehat{\mathbf{r}}%
_{1}\right)  =\overline{K}_{1}\left(  R\right)  $, $K_{2}\left(
\widehat{\mathbf{r}}_{1},\widehat{\mathbf{r}}_{2}\right)  =\overline{K}%
_{2}\left(  \widehat{\mathbf{r}}_{1}\cdot\widehat{\mathbf{r}}_{2};R\right)
,$etc. Substituting the 0D densities and making use of the stability property
of the kernels, one finds that
\begin{equation}
F_{\text{ex}}\left[  \rho_{1}\right]  =F_{\text{ex}}^{\left(  1\right)
}\left[  \rho_{1}\right]  =\overline{K}_{1}\psi_{0}\left(  N_{1}\right)
\frac{\partial}{\partial R}V_{D}\left(  R\right)
\end{equation}
where $V_{D}\left(  R\right)  $ is the volume of a sphere of radius $R$ in $D$
dimensions and where $\psi_{1}\left(  x\right)  =\frac{d}{dx}\psi_{0}\left(
x\right)  \equiv\psi_{0}^{\prime}\left(  x\right)  $. (See Appendix \ref{appB} for the derivation of this and similar results discussed in this Section.) This reproduces the
exact result provided $\psi_{0}\left(  x\right)  =\Phi_{0}\left(  x\right)  $
and $\overline{K}_{1}\frac{\partial}{\partial R}V_{D}\left(  R\right)
=\overline{K}_{1}S_{D}\left(  R\right)  =1$, where $S_{D}\left(  R\right)  $
is the surface area of the sphere.

For two cavities, after identifying $\psi_{2}\left(  x\right)  =\Phi
_{0}^{\prime\prime}\left(  x\right)  $, one has that
\begin{align}
F_{\text{ex}}\left[  \rho_{2}\right]   &  =F_{\text{ex}}^{\left(  1\right)
}\left[  \rho_{2}\right]  +F_{\text{ex}}^{\left(  2\right)  }\left[  \rho
_{2}\right] \\
&  =\Phi_{0}\left(  N_{1}+N_{2}\right)  +\left(  \Phi_{0}\left(  N_{1}%
+N_{2}\right)  -\Phi_{0}\left(  N_{1}\right)  -\Phi_{0}\left(  N_{2}\right)
\right)  \left\{  \Delta_{2}\left(  s_{12}\right)  -1\right\} \nonumber
\end{align}
where $\Delta_{2}\left(  s_{12}\right)  $ vanishes if the two cavities do not
intersect (thus giving the exact result for this case) and otherwise, in 3
dimensions, it is
\begin{equation}
\Delta_{2}\left(  s_{12}\right)  =1-\frac{s_{12}}{2R}+\frac{4\pi R^{2}}%
{s_{12}}\overline{K}_{2}\left(  1-\frac{s_{12}^{2}}{2R^{2}};R\right)  .
\end{equation}
The generalization to any number of dimensions is given in Eq.(\ref{B16}) of the Appendix. In
order to reproduce the exact result, one needs that $\Delta_{2}\left(
s_{12}\right)  =1$ so
\begin{equation}
\overline{K}_{2}\left(  1-\frac{s_{12}^{2}}{2R^{2}};R\right)  =\frac{1}{4\pi
R}\frac{s_{12}^{2}}{2R^{2}}%
\end{equation}
or, equivalently,
\begin{equation}
\overline{K}_{2}\left(  x;R\right)  =\frac{1}{4\pi R}\left(  1-x\right)
\end{equation}

Before proceeding, it is interesting to see the correspondence between this
form of FMT and the more familiar standard form above. Comparing the two, one
sees that if $\overline{K}_{n}\left(  x_{1},x_{2},...;R\right)  $ is a
polynomial of the form $a_{0}+a_{1}\left(  \widehat{\mathbf{r}}_{1}%
\cdot\widehat{\mathbf{r}}_{2}\right)  +a_{11}\left(  \widehat{\mathbf{r}}%
_{1}\cdot\widehat{\mathbf{r}}_{2}\right)  ^{2}+a_{123}\left(  \widehat
{\mathbf{r}}_{1}\cdot\widehat{\mathbf{r}}_{2}\right)  \left(  \widehat
{\mathbf{r}}_{2}\cdot\widehat{\mathbf{r}}_{3}\right)  +...$etc then
$F_{\text{ex}}^{\left(  n\right)  }\left[  \rho\right]  $ can be written as
\begin{equation}
F_{\text{ex}}^{\left(  n\right)  }\left[  \rho\right]  =\int\psi_{n}\left(
\eta\left(  \mathbf{r}\right)  \right)  \left\{
\begin{array}
[c]{c}%
a_{0}s^{n}\left(  \mathbf{r}\right)  +a_{1}s^{n-2}\left(  \mathbf{r}\right)
v^{2}\left(  \mathbf{r}\right) \\
+a_{11}s^{n-2}\left(  \mathbf{r}\right)  \operatorname*{Tr}T^{2}\left(
\mathbf{r}\right)  +a_{123}s^{n-3}\left(  \mathbf{r}\right)  \mathbf{v}\left(
\mathbf{r}\right)  \cdot\mathbf{T}\left(  \mathbf{r}\right)  \cdot
\mathbf{v}\left(  \mathbf{r}\right)  ...
\end{array}
\right\}  d\mathbf{r}%
\end{equation}
so that in particular with the kernels given above,
\begin{equation}
F_{\text{ex}}^{\left(  1\right)  }\left[  \rho_{2}\right]  +F_{\text{ex}%
}^{\left(  2\right)  }\left[  \rho_{2}\right]  =\int\left\{  \frac{1}{4\pi
R^{2}}s\left(  \mathbf{r}\right)  \ln\left(  1-\eta\left(  \mathbf{r}\right)
\right)  +\frac{1}{4\pi R}\frac{s^{2}\left(  \mathbf{r}\right)  -v^{2}\left(
\mathbf{r}\right)  }{\left(  1-\eta\left(  \mathbf{r}\right)  \right)
}\right\}  d\mathbf{r}%
\end{equation}
which are the first two terms of Rosenfeld's functional in three dimensions.
One sees that it is only possible to write the free energy functional in the
standard FMT formulation if the kernels are polynomial functions of their arguments.

To describe three cavities, let $V_{ij}$ be the volume of intersection of
spheres of radii $R_{i}$ and $R_{j}$ with centers $\mathbf{s}_{i}$ and
$\mathbf{s}_{j}$ and let $V_{ijk}$ be the volume of mutual intersection for
three such spheres. Then, for $V_{12}=V_{13}=V_{23}=0$ (case (a) in Fig.
\ref{fig0}) the exact result is $F_{\text{ex}}\left[  \rho_{3}\right]
=\Phi_{0}\left(  N_{1}\right)  +\Phi_{0}\left(  N_{2}\right)  +\Phi_{0}\left(
N_{3}\right)  $ and this is reproduced by the FMT ansatz, provided the kernels
satisfy the stability condition. For $V_{12}\neq0,V_{13}=V_{23}=0$ (case (b))
both give $F_{\text{ex}}\left[  \rho_{3}\right]  =\Phi_{0}\left(  N_{1}%
+N_{2}\right)  +\Phi_{0}\left(  N_{3}\right)  $. The cases $V_{12}\neq
0,V_{13}\neq0,V_{23}=0$ (case (c) in the Figure) and $V_{12}\neq0,V_{13}%
\neq0,V_{23}\neq0,V_{123}=0$ (case (d)) give rise to more complicated exact
results while FMT gives $\Phi_{0}\left(  N_{1}+N_{2}\right)  +\Phi_{0}\left(
N_{1}+N_{3}\right)  -\Phi_{0}\left(  N_{1}\right)  $ in the first case and
$\Phi_{0}\left(  N_{1}+N_{2}\right)  +\Phi_{0}\left(  N_{1}+N_{3}\right)
+\Phi_{0}\left(  N_{2}+N_{3}\right)  -\Phi_{0}\left(  N_{1}\right)  -\Phi
_{0}\left(  N_{2}\right)  -\Phi_{0}\left(  N_{3}\right)  $ in the second case,
neither of which is correct. Furthermore, there is clearly no 
freedom within this ansatz to capture the correct results. Rosenfeld and
Tarazona label these "lost cases" as they simply cannot be described within
this framework. The final possibility, $V_{123}\neq0$ (case (e) in Fig.
\ref{fig0}) yields the expected exact result $\Phi_{0}\left(  N_{1}%
+N_{2}+N_{3}\right)  $ and the ansatz gives, after identifying $\psi
_{3}\left(  x\right)  =\Phi_{0}^{\prime\prime\prime}\left(  x\right)  $,
\begin{align}
&  F_{\text{ex}}^{\left(  1\right)  }\left[  \rho_{2}\right]  +F_{\text{ex}%
}^{\left(  2\right)  }\left[  \rho_{2}\right]  +F_{\text{ex}}^{\left(
3\right)  }\left[  \rho_{2}\right] \\
&  =\Phi_{0}\left(  N_{1}+N_{2}+N_{3}\right)  +\left\{  \sum_{i}\Phi
_{0}\left(  N_{i}\right)  \mathbf{-}\sum_{i<j}\Phi_{0}\left(  N_{i}%
+N_{j}\right)  \mathbf{+}\Phi_{0}\left(  N\right)  \right\}  \left\{
\Delta_{3}\left(  s_{12},s_{13},s_{23}\right)  -1\right\} \nonumber
\end{align}
where $\Delta_{3}\left(  s_{12},s_{13},s_{23}\right)  $ vanishes if
$V_{123}=0$ and otherwise it is
\begin{equation}
\Delta_{3}\left(  s_{12},s_{13},s_{23}\right)  =\lim_{R_{1},R_{2}%
,R_{3}\rightarrow R}\left\{
\begin{array}
[c]{c}%
\overline{K}_{1}\sum_{i=1}^{3}\frac{\partial}{\partial R_{i}}V_{123}+\frac
{1}{4\pi R}\sum_{i=1}^{3}\sum_{j=i+1}^{3}\frac{s_{ij}^{2}}{R^{2}}%
\frac{\partial^{2}}{\partial R_{i}\partial R_{j}}V_{123}\\
+6\overline{K}_{3}\left(  ...\right)  \frac{\partial^{3}}{\partial
R_{1}\partial R_{2}\partial R_{3}}V_{123}%
\end{array}
\right\}
\end{equation}
where the arguments of $\overline{K}_{3}$ have been suppressed for clarity.
Reproduction of the exact result demands that $\Delta_{3} = 1$ thus determining $\overline{K}_{3}$.
The various volume derivatives can be worked out but they are quite
complicated, involving square roots, inverse trigonometric functions and many
different cases:\ the exact expression can clearly not be written as a
polynomial, except perhaps as an infinite expansion. For example, for the most
symmetric case $s_{12}=s_{13}=s_{23}=D$, it simplifies to
\begin{equation}
\overline{K}_{3}\left(  x,x,x\right)  =\frac{1}{48}\left(  1-x\right)
\sqrt{1+2x}\left(  1-\frac{6}{\pi}\arcsin\frac{x}{1+x}\right)
\end{equation}
which we can confirm vanishes at $x=1$, as demanded by the stability condition.

In summary, the dimensional crossover program has successfully generated the
first and second contributions to the excess free energy functional of
Rosenfeld's FMT. It is known that the sum of these two terms also reproduces
the exact 1D functional (the Percus functional) when the density is suitably
restricted and it is easy to show that the higher order terms do not
contribute to this limit provided they satisfy the stability condition (see
Appendix \ref{1D}). However, for three cavities, only the simplest cases of at
least one completely disjoint sphere are reproduced correctly (and these cases
do not involve $K_{3}$). Two classes of configurations are a priori
inaccessible to the ansatz while the third, involving three cavities that have
a non-zero mutual intersection, can be recovered but the kernel is a
complicated, non-polynomial function of its arguments. The latter fact means
that it cannot be expressed in terms of a finite number of fundamental
measures. From a practical point of view, it is imperative to have such a
formulation in terms of fundamental measures expressed as convolutions that they can be evaluated efficiently. This impasse suggests that we
ask what properties could be preserved by an "acceptable"
kernel. Evidently, it must be a polynomial function of its arguments and if we
want to restrict the possibilities to those that generate functionals
involving only the fundamental measures already discussed, then the most
general form possible, taking account of symmetry, is
\begin{equation}
\overline{K}_{3}\left(  x,y,z\right)  =a+b\left(  x+y+z\right)  +c(x^{2}%
+y^{2}+z^{2})+d\left(  xy+xz+yz\right)  +exyz
\end{equation}
The stability condition demands that $\overline{K}_{3}\left(  x,x,1\right)  $
vanish, so
\begin{equation}
0=\left(  a+b+c\right)  +\left(  2b+2d\right)  \allowbreak x+\left(
2c+d+e\right)  x^{2}%
\end{equation}
$\allowbreak$and since $x$ can vary freely, the various coefficient must
vanish leaving, after renaming the remaining constants,
\[
\overline{K}_{3}\left(  x,y,z\right)  =\frac{1}{24\pi}A\left(  1-x\right)
\left(  1-y\right)  \left(  1-z\right)  +\frac{1}{24\pi}B\allowbreak\left(
1-x^{2}-y^{2}-z^{2}+2xyz\right)
\]
which, in terms of the vectors, can be written as
\begin{equation}
K_{3}\left(  \widehat{\mathbf{r}}_{1},\widehat{\mathbf{r}}_{2},\widehat
{\mathbf{r}}_{3}\right)  =\frac{1}{24\pi}A\left(  1-\widehat{\mathbf{r}}%
_{1}\cdot\widehat{\mathbf{r}}_{2}\right)  \left(  1-\widehat{\mathbf{r}}%
_{1}\cdot\widehat{\mathbf{r}}_{3}\right)  \left(  1-\widehat{\mathbf{r}}%
_{2}\cdot\widehat{\mathbf{r}}_{3}\right)  +\frac{1}{24\pi}B\allowbreak
\left\vert \widehat{\mathbf{r}}_{1}\cdot\left(  \widehat{\mathbf{r}}_{2}%
\times\widehat{\mathbf{r}}_{3}\right)  \right\vert ^{2}.
\end{equation}
Thus, the only practical question is how to determine the constants, $A$ and
$B$. The original FMT of Rosenfeld is recovered by taking $A=9/8,B=0$ and
keeping only the lowest order (linear) terms involving the scalar products
while the tensor version of Tarazona corresponds to $A=-B=3/2.$ Note that the
coefficients of the two constants are obviously non-negative so that
$A,B\geq0$ automatically assures stability of the resulting functional.
Indeed, since $\overline{K}_{3}\left(  -0.5,-0.5,-0.5\right)  =\frac{9}{64\pi
}A$ the kernel can only be non-negative for all arguments if $A\geq0$.
Similarly, $\overline{K}_{3}\left(  1-\varepsilon,1-\varepsilon,1-\varepsilon
\right)  =\frac{1}{24\pi}\varepsilon^{2}\left(  \varepsilon A+\left(
3-2\varepsilon\right)  B\right)  $ so non-negativity also demands for
vanishingly small $\varepsilon$ also demands that $B\geq0$. The fact that the
existing tensor functionals are well outside these limits is most likely the
source of their instability. In the following, the class of theories discussed here with coefficients $A,B>0$ will be called ``explicitly stable FMT'' or ``esFMT(A,B)''. The the qualification is due to the fact that it is possible that negative coefficients could also yield a stable (in the sense of bounded from below) theory but this is not obvious. 

\section{Properties of explicitly stable functionals}

In summary, the class of "explicitly" stable functionals consistent with
dimensional crossover and involving only measures up to the second-order
tensor are $F_{\text{ex}}\left[  \rho\right]  =F_{\text{ex}}^{\left(
1\right)  }\left[  \rho\right]  +F_{\text{ex}}^{\left(  2\right)  }\left[
\rho\right]  +F_{\text{ex}}^{\left(  3\right)  }\left[  \rho\right]  $ with
the third contribution having the form%
\begin{equation}
F_{\text{ex}}^{\left(  3\right)  }\left[  \rho\right]  =\frac{1}{24\pi}%
\int\frac{1}{\left(  1-\eta\left(  \mathbf{r}\right)  \right)  ^{2}}\left\{
\begin{array}
[c]{c}%
\left(  A+B\right)  s^{3}\left(  \mathbf{r}\right)  -3As\left(  \mathbf{r}%
\right)  v^{2}\left(  \mathbf{r}\right)  +3A\mathbf{v\left(  \mathbf{r}%
\right)  \cdot T\left(  \mathbf{r}\right)  \cdot v}\left(  \mathbf{r}\right)
\\
-3Bs\left(  \mathbf{r}\right)  \operatorname*{Tr}\mathbf{T}^{2}\left(
\mathbf{r}\right)  +\left(  2B-A\right)  \operatorname*{Tr}\mathbf{T}%
^{3}\left(  \mathbf{r}\right)
\end{array}
\right\}  d\mathbf{r}%
\end{equation}
with $A,B\geq0$. The original proposal of Tarazona is not in this class as it
requires $A=-B=\frac{3}{2}$. As shown below, any combination $A+B/4=1$ gives the Percus-Yevik (PY)
equation of state but the PY direct correlation function can only be obtained
the Tarazona model (which, indeed, was the reason for its form) . These "explicitly stable" models give finite results for any of the combinations of 0D cavities and are
positive semi-definite (hence implying that the free energy is bounded from
below for any density field). Having defined, as precisely as possible, the
class of explicitly-stable models, this section is devoted to investigating
the results of various choices for the constants $A$ and $B$.

The next subsection deals with the properties of the homogeneous fluid, the
following with the inhomogeneous fluid and the remainder with the solid phase.
Calculations on inhomogeneous systems were performed by discretizing the
density field on a cubic lattice with spacing $\Delta$ which is typically much
smaller than a hard-sphere radius. Details of the calculations (using analytic
representations of the fundamental measures and a consistent real-space scheme
to evaluate the convolutions) have been given in a recent
publication\cite{Schoonen}. This discretized density is used to evaluate the
functional $\Lambda[\rho]$ which is then minimized with respect to the density
in a scheme we refer to as ``full minimization'' since, beyond the
discretization, there are no constraints on the calculations. For the solid
phases, some calculations were also performed using an approximate
``constrained'' scheme whereby the density is written as a sum of Gaussians,
\begin{equation}
\rho(\mathbf{r}) = (1-c)\sum_{\mathbf{R}}\left(  \frac{\alpha}{\pi} \right)
^{3/2}exp(-\alpha(\mathbf{r}-\mathbf{R})^{2})
\end{equation}
where $0 \le c <1$ is the vacancy concentration, the sum is over the Bravais
lattice sites and the parameter $\alpha$ controls the width of the Gaussians.
For a face-centered cubic (FCC) or body-centered cubic (BCC) lattice, this
represents a solid with lattice density $\rho_{\text{latt}} = 4/a^{3}$ or
$2/a^{3}$ respectively where $a$ is the cubic lattice constant. In both cases,
the average number density (e.g. the integral of the density over all space
divided by the volume) is $\bar{\rho} = (1-c)\rho_{\text{latt}}$. By
converting this sum to Fourier-space it is easy to see that $\alpha=0$
corresponds to a uniform density of $\bar{\rho}$ so that varying $\alpha$
allows one to go smoothly from the fluid to the solid state. In our
Gaussian calculations, the vacancy concentration is generally fixed to some small
value, e.g. $c = 10^{-4}$, and we minimize with respect to $\alpha$.
Generally, most properties of the solid are insensitive to the precise value
of $c$ but one could of course minimize with respect to this parameter as well. Note that in general, the vacancy concentration is calculated as $c=(N_{latt}-N[\rho])/N_{latt}$ where $N_{latt}$ is the number of lattice sites in the system (i.e. 4 per cubic unit cell for FCC and 2 for BCC). Finally, a standard measure of the widths of the distributions is the Lindemann parameter defined as the root mean square displacement divided by the nearest neighbor distance. 

\subsection{Properties the homogeneous fluid}

Keeping the constants arbitrary for the moment, the resulting equation of
state for the homogeneous fluid ($\rho\left(  \mathbf{r}\right)
\rightarrow\rho$ and $\eta\left(  \mathbf{r}\right)  \rightarrow\eta=\pi
\rho\sigma^{3}/6$) is
\begin{equation}
\label{eos}\frac{\beta P}{\rho}\equiv Z\left(  A,B\right)  =Z^{PY}+\left(
\frac{2}{3}\left(  4A+B\right)  -3\right)  \frac{\eta^{2}}{\left(
1-\eta\right)  ^{3}}%
\end{equation}
where the Percus-Yevik compressibilty factor is
\begin{equation}
Z^{PY}=\frac{1+\eta+\eta^{2}}{\left(  1-\eta\right)  ^{3}}%
\end{equation}
and we recall that both the Rosenfeld and Tarazona functionals reproduce PY.
The first several terms of the virial expansion are given in Table \ref{tab1}.  The direct correlation function for the homogeneous liquid is calculated from the exact relation\cite{lutsko:acp} $c(\mathbf{r}_1,\mathbf{r}_{2};[\rho]) = -\delta^2 \beta F_{ex}[\rho]/\delta \rho(\mathbf{r}_1) \delta \rho(\mathbf{r}_2)$ evaluated at constant density yielding%
\begin{equation}
c\left(  r=\sigma x;A,B\right)  =c^{PY}\left(  r\right)  +\left(
\begin{array}
[c]{c}%
-\frac{1}{2}\frac{\eta}{\left(  1-\eta\right)  ^{2}}x\left(  2\left(
A+B\right)  x^{2}+3-2B-4A\right) \\
+\left(  8A+2B-9\right)  \frac{\eta^{2}}{\left(  1-\eta\right)  ^{3}}\left(
1-x\right)  \left(  1+\frac{1}{2}\frac{\eta}{1-\eta}\left(  1-x\right)
\left(  2+x\right)  \right)  \;
\end{array}
\right)  \Theta\left(  1-x\right)
\end{equation}
To recover the PY equation of state requires that $4A+B=\frac{9}{2}$ in which
case the PY direct correlation function is also recovered with $A=-B$, which
is again Tarazona's model.

\begin{table}
\begin{minipage}{\linewidth}
\caption{The nth virial coefficient in the PY approximation, the result of esFMT for arbitrary values of the coefficients $A,B$ and where $C =\frac{1}{3}\left(8A+2B-9\right)$ and the exact values\cite{Virial}. }
\label{tab1}
\addtolength\tabcolsep{2pt}
\begin{tabular}[c]{cccc}%
\hline \hline
$n$ & $B_{n}^{\left(  \text{PY}\right)  }$  & $B_{n}^{(\text{esFMT})}$ & $B_{n}^{\left(  \text{exact}\right)  }$\\
\hline
2 & 4 & 4 & 4 \\
3 & 10 & 10+C & 10 \\
4 & 19 & 19+3C & 18.36 \\
5 & 31 & 31+6C & 28.22 \\
6 & 46 & 46+10C & 39.82 \\
7 & 64 & 64+15C & 53.34 \\
\hline
\hline
\end{tabular}
\end{minipage}
\end{table}

\begin{figure}
[htp!]
\includegraphics[width=0.8\linewidth]{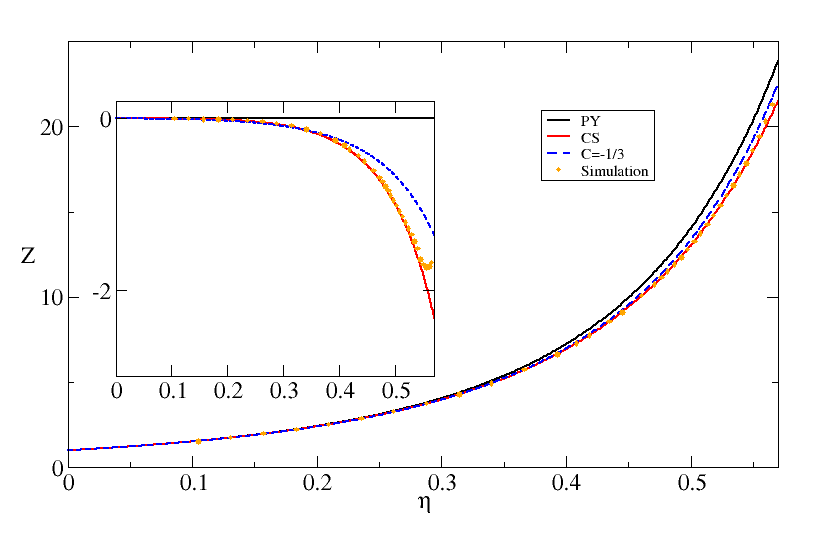}
\caption{Compressibility factor $Z=\beta P/\rho$ for the homogeneous liquid as a function of density as predicted by the Percus-Yevik (PY) and Carnahan-Starling (CS) equations of state, Eq.(\ref{eos}) with $C=-1/3$ and simulation data reported in Ref.\cite{Zsimulation}. The inset shows the difference of the various quantities from the PY prediction. }
\label{fig1}
\end{figure}

\begin{figure}
[htp!]\includegraphics[width=0.80\linewidth]{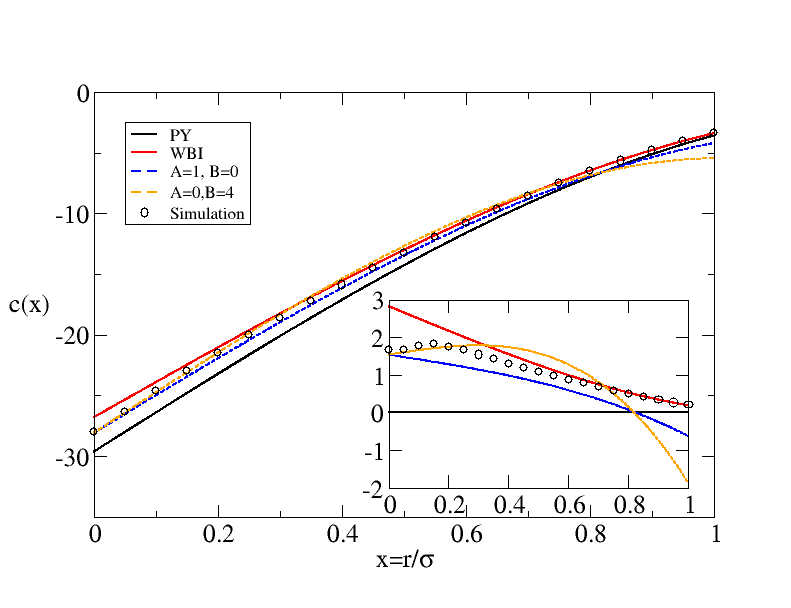}
\caption{Direct correlation function for the uniform hard-sphere fluid at density $\rho\sigma^3 = 0.8$, or $\eta = 0.4189$, as predicted by the Percus-Yevik theory, the WBI model, the esFMT with two different choices of parameters (which give the same equation of state) and simulation data reported in Ref. \cite{Alder_Young_DCF}. The inset shows the differences from the PY result. Note that the exact DCF is also non-zero outside the hard-core but none of the theories considered here allow for this. }
\label{fig2}
\end{figure}

So, if one wishes to be sure that the resulting functional is stable for any
density distribution, and therefore demands that $A,B\geq0$, then it is
impossible to reproduce the PY direct correlation function. In any case, this
not an exact result so there is no particular requirement to do so (and in
fact, the White Bear functionals do not). However, since PY is exact up to
second order in the density, a difference from the PY expression of first
order in the density also means that one does not recover the exact dcf up to
first order which may be more disturbing but, unavoidable, if the functional
is to be explicitly stable. A reasonable result is achieved if one takes, e.g.
for simplicity, $A=1$ and $B=0$ which means that $C=-\frac{1}{3}$. This
creates a small error at order $\eta^{2}$ but improves the virial
expansion at higher order. For example, it gives $B_{2}=9\frac{2}{3}$ and
$B_{7}=59$ compared to the exact values of $10$ and $53.34$, respectively and the compressibility factor is generally improved by this choice
relative to PY, see Fig. \ref{fig1}. Figure \ref{fig2} shows the dcf for
$(A,B)=(1,0)$ and $(0,4)$, both corresponding to $C=-\frac{1}{3}$, compared to
PY, the dcf generated by the WBI approximation and simulation data. At small
values of $r$, the dcf generated by the stable FMT is actually closer to
simulation than that of WB or the PY result. At the largest value, $r=\sigma$,
the esFMT is not as close but the deviations are not very large.

\subsection{Fluid near a wall}

\begin{figure}
[htp!]\includegraphics[width=0.80\linewidth]{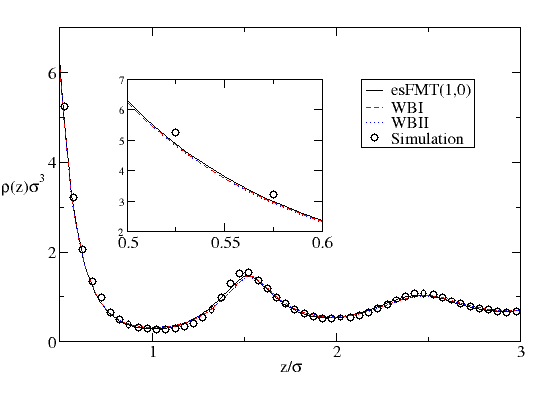}
\caption{Fluid structure near a wall with bulk density $\rho \sigma^3 = 0.813$, or $\eta = 0.426$,  for the White Bear models and the esFMT(1,0) model. All calculations were performed on a grid of $1 \times 1 \times 10,000$ points with spacing $\Delta = 0.01\sigma$. The simulation results are from Ref.\cite{Groot}. The inset shows the slight differences at the point of contact with the wall.}
\label{wall1}
\end{figure}

\begin{figure}
[htp!]\includegraphics[width=0.80\linewidth]{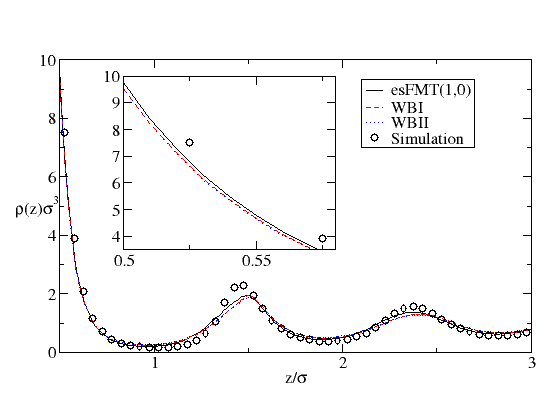}
\caption{As in Fig.\ref{wall1} for a density of $\rho \sigma^3 = 0.9135$, or $\eta = 0.478$.}
\label{wall2}
\end{figure}

As a first example of an inhomogeneous system, the structure of a fluid in
contact with a wall is shown in Figs.\ref{wall1} and \ref{wall2} for bulk
densities $\rho\sigma^{3}=0.813$ and $0.9135$ corresponding to packing
fractions of $0.426$ and $0.478$ respectively. The esFMT agrees quite closely
with the WBI and WBII functionals and all are in reasonable agreement with
simulation at the lower density but show similar deviations at the higher
density. This is due to the fact that all FMT's obey the
wall-theorem\cite{lutsko:acp} which says that the density of a fluid of hard
spheres at the point of contact with a hard wall will be $\beta P$ so, since
the esFMT equation of state is somewhat different than that of the WB models,
the contact density necessarily differs also. In any case, the differences
between all of these models is quite small.

\subsection{Freezing transition}

The freezing of hard-spheres (into an FCC solid) has long served as a
benchmark problem for the development of new model functionals. In the
framework of cDFT the procedure is conceptually straightforward: one must
minimize the density functional at fixed chemical potential (i.e in the grand
canonical ensemble) to find the meta-stable states: typically, one constant
density (fluid) state and a solid state with the density localized at the
Bravais lattice sites. The state with the smallest free energy is
thermodynamically favored and at some chemical potential, the free energies of
the two states are equal thus determining the freezing transition.

In early work, the density was almost always modeled as a sum of Gaussians (or
sometimes, distorted Gaussians) centered at the FCC lattice sites. In the
simplest case, the Gaussians have a fixed normalization (less than or equal to
one) and the free energy functional reduces to a function of two parameters,
the width of the Gaussians and the FCC lattice constant. At little more
expense, one can allow the normalization of the Gaussians to vary as well
yielding a function of three parameters. In both cases, the function must then
be minimized to determine the equilibrium state. It happens that Gaussians
with infinite width give a uniform density which corresponds to the
homogeneous fluid so by varying the width, one passes from the fluid to the
solid thus allowing both to be described in the same framework. More recent
applications (including the present work) simply discretize the density on a
grid (with computational unit cells much smaller than the FCC unit cell) and
one then minimizes with no constraints. These calculations are usually
performed in two steps: minimization at constant lattice parameter, and then
minimization over the lattice parameters. Further details of the implementation
used in this work can be found in \cite{Schoonen}. Table \ref{tab2} shows the
results obtained for the freezing transition using the esFMT, the original
Tarazona theory (corresponding to $A=-B=3/2$), the WBI and WBII theories and
simulation results.

\begin{table}
\begin{minipage}[v]{\linewidth} 
\caption{Freezing parameters as calculated using several different FMT models and simulation results. All calculations were performed using full minimization on a lattice with spacing $\Delta = 0.025\sigma$ except for the Tarazona model which was estimated using Gaussian profiles. The Lindemann parameter is the square root of the mean-squared displacement divided by the nearest-neighbor distance. The simulation results are from Fortini and Dijkstra\cite{Fortini_2006}, except the Lindemann parameter taken from \cite{Alder_Young_DCF} and the vacancy concentration which is an estimate based on Ref.\cite{Bennett}. For the esFMT, calculations on a lattice with spacing $\Delta = 0.0125\sigma$ yield virtually identical results. The table also includes results based on a more heuristic approach to explicit stability called ``mRSLT'' in Ref.\cite{LutskoLam}.}
\label{tab2}
\addtolength\tabcolsep{2pt}
\begin{tabular}[c]{ccccccc}%
\hline \hline
Source & $\eta_{liq}$  & $\eta_{solid}$ & $\beta P \sigma^3$ & $\beta \mu$ & Lindemann Parameter & Vacancy Concentration\\
\hline
WBI & $0.491$ & $0.534$ & $11.50$ & $16.0$ & $0.138$ &  $-7 \times 10^{-6}$ \\
WBII & $0.498$ & $0.544$ & $12.17$ & $16.70$ & $0.122$ &  $4 \times 10^{-5}$ \\
Tarazona  & $0.472$ & $0.518$ & $10.04$ & $14.56$  & $0.149$ & --- \\
esFMT(1,0)  & $0.486$ & $0.533$ & $11.28$ & $15.8$  & $0.141$ & $6 \times 10^{-4}$ \\
mRSLT & $0.514$ & $0.547$ & $14.14$ & $18.73$ & $0.133$ & $-2 \times 10^{-4}$ \\
\hline
Simulation & $0.4915(5)$ & $0.5428(5)$ & $11.57(10)$ & $16.08(10)$ & $0.126$ & $1.4 \times 10^{-4}$ \\
\hline
\hline
\end{tabular}
\end{minipage}

\end{table}

The explicitly stable model with $A=1$, $B=0$ gives results quite similar to
WBI: this may be a little surprising since the quality of the prediction of
freezing is tied to the accuracy of the fluid equation of state and the
Carnahan-Starling equation of state that is built into WBI is better than that
of esFMT(1,0) . In any case, as for the homogeneous fluid, the
esFMT seems to perform almost as well as the state of the art. In particular,
Table 2 includes results for a non-tensorial model that is also positive
semi-definite, but which does not correctly describe multiple 0D cavities (the model discussed in Ref.\cite{LutskoLam} and referred to as ``mRSLT'').
Despite incorporating the same equation of state as WBI, it performs much
worse in predicting hard-sphere freezing.

Finally, the table also gives the vacancy concentration at freezing
(calculated for a unit cell of the FCC lattice as $c = \frac{4-N}{V}$ with $N$
the total number of molecules in the cell and $V$ its volume). These numbers
are not necessarily very accurate since the grid spacing ($\Delta=
0.025\sigma$) does not allow for a very accurate representation of the density
peaks when they are too sharp but it is consistent with calculations on finer
grids. Given that the values determined from simulation for densities near
freezing are on the order of $10^{-4}$ the important point is that, as noted
previously\cite{LutskoLam}, the WB models give typical values about an order
of magnitude smaller than simulation while the esFMT seems to improve on this somewhat.

\subsection{Solid structure and thermodynamics}

\begin{figure}
[htp!]\includegraphics[width=0.80\linewidth]{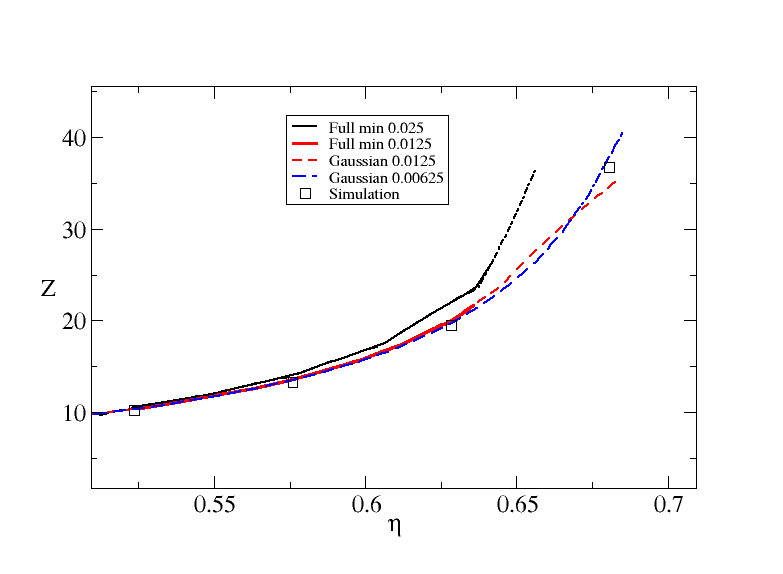}
\caption{Pressure factor as determined from the esFMT(1,0) theory using full minimization with a computational lattice spacing of $\Delta = 0.025\sigma$ and minimization of Gaussians with spacings $\Delta = 0.025\sigma$, $0.0125\sigma$ and $0.00625\sigma$. Simulation results are taken from  Bannerman et al.\cite{HardSpherePressure}. The results of full minimization and of the Gaussian approximation are indistinguishable.}
\label{Z}
\end{figure}

Figure \ref{Z} shows the pressure factor for the FCC solid as a function of
packing fraction for the esFMT(1,0) model compared to section. The
calculations were performed using full minimization with spacings  $\Delta=
0.025\sigma$ and $0.0125\sigma$ as well as minimization of Gaussian profiles on this and finer
lattices. Little difference is seen between the full minimization and the
Gaussian model. In each case, the results are in good agreement with
simulation up to some maximum packing fraction at which point the solid peaks
become so narrow that that the lattice is no longer able to accurately
represent them.

\begin{figure}
[htp!]\includegraphics[width=0.80\linewidth]{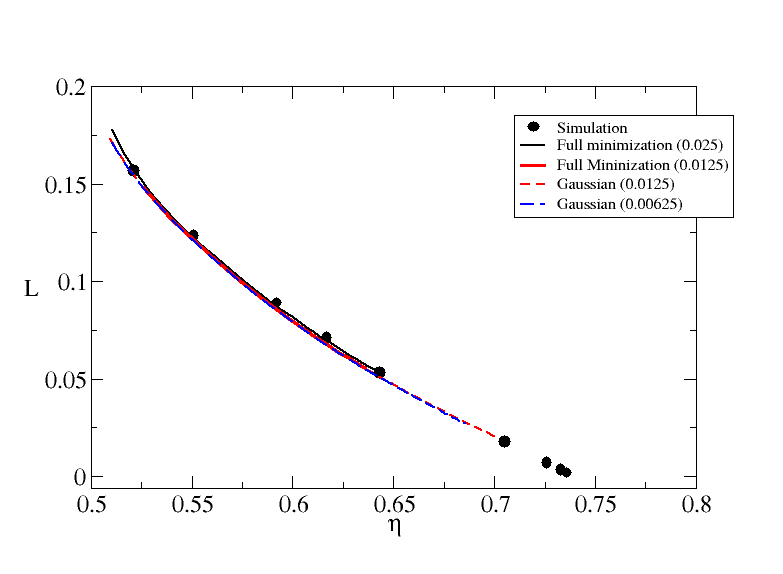}
\caption{Same as Fig. \ref{Z} but showing the Lindemann parameter. The simulation results are from Alder and Young\cite{Alder_Young_DCF}. Again, the constrained and unconstrained results are virtually identical.}
\label{L}
\end{figure}

Figure \ref{L} shows a similar comparison for the ratio of the
root-mean-squared displacement divided by the nearest neighbor distance, i.e.
the Lindemann parameter. Again, all of the calculations are in good agreement
with simulation up to the limits imposed by the lattice.

\begin{figure}
[htp!]\includegraphics[width=0.80\linewidth]{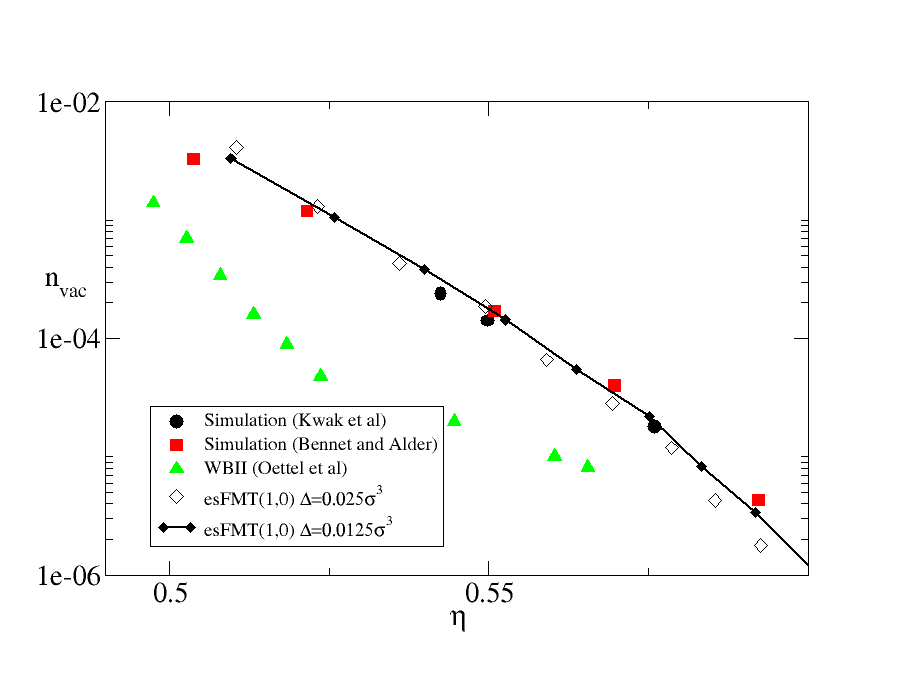}
\caption{The vacancy concentration as a function of density as determined by full minimization of esFMT(1,0) with lattice spacing $\Delta = 0.0125\sigma$ and $0.00625\sigma$, the results reported by Oettel et al\cite{Oettel2010}, and from simulations by Kwak et al\cite{Kwak} and by Bennet and Alder\cite{Bennett}.}
\label{V}
\end{figure}

The vacancy concentration is shown in Fig.\ref{V}. It is a very small quantity
compared to the density and so is quite sensitive to numerical noise but the
values obtained from the esFMT(1,0) model are again in reasonable agreement
with simulation.

\subsection{The BCC crystal}

One of the weaknesses of FMT is the description of non-FCC crystals. Older
forms of cDFT never gave very good descriptions of the BCC structure at high
density: in particular, the Lindemann parameter only decreased slightly as
density increased and eventually began to increase with increasing density,
which is very unphysical behavior(see, e.g., \cite{Gela}). One of the
impressive accomplishments of the original tensor version of FMT is that it
predicts that the Lindemann parameter of the BCC crystal goes to zero at BCC
close packing just as it should\cite{Tarazona}. However upon further
examination\cite{Lutsko_FCC}, it was found that this was preceded by a
similarly unphysical behavior and that in fact that there are really two BCC
structures: one that shows the same unphysical behavior as the older theories
gave and one that gives the new, more physical behavior. Furthermore, the
unphysical structure has the lower free energy for packing fractions below
about $\eta= 0.6$. The White Bear theory was even worse with the unphysical
structure always being the more stable of the two. So it is interesting to ask
whether the esFMT does any better or worse.

\begin{figure}
[htp!]\includegraphics[width=0.80\linewidth]{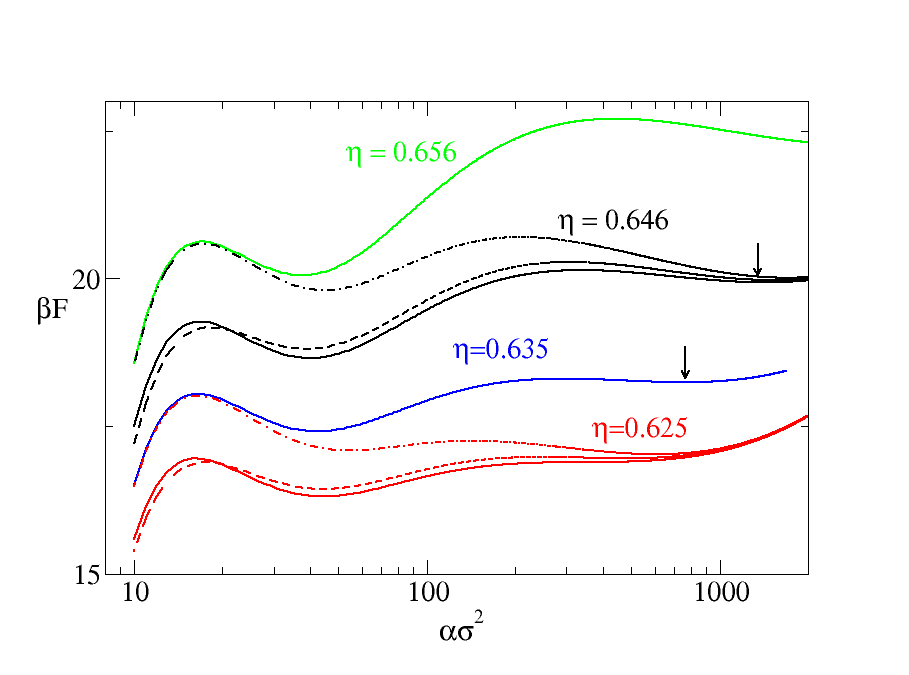}
\caption{Free energy divided by temperature  as a function of the Gaussian parameter, $\alpha$, for the BCC crystal at several different densities as determined using the esFMT(1,0) theory. The minima, marked with arrows, correspond to metastable crystal states for the esFMT(1,0) model. The dashed-lines are the WBI results and the dot-dash lines are from Tarazona's original tensor model at $\eta = 0.625$ and $0.646$.}
\label{bcc1}
\end{figure}

In Fig. \ref{bcc1} the free energy of the BCC structure calculated in the
Gaussian approximations is shown as a function of the Gaussian parameter
$\alpha$ for several different densities. The liquid state corresponds to
$\alpha= 0$ and one sees that there is indeed a solid-like minimum as well but
at a value that is relatively low compared to the FCC solid. This is not
surprising since the BCC structure is not close-packed and so the molecules
are expected to have more room to move. However, it is apparent from the
figure that this minimum does not really increase with increasing density and
that at high densities, a second minimum appears at much higher values of
$\alpha$. This second minimum does increase quickly with density near close
packing as one would expect and in all cases, it is very shallow leaving the
low-$\alpha$ minimum as the thermodynamically more stable of the two. This
behavior is very similar to that of the WBI theory and shows that, as in
other forms of FMT, the stable theory does not give a particularly physical
description of the BCC state.

The figure also shows results using WBI and Tarazona's original tensor model.
The similarity between esFMT(1,0) and WBI is again evident with the curves
being nearly indistinguishable. This is not the case with Tarazona's model
which shows a considerably higher first barrier and minimum than the other two
whereas the second minimum is quite similar in position to that of the the
other two theories. Recall that Tarazona's model gives the PY equation of
state for the homogeneous fluid (corresponding to $\alpha= 0$) whereas WBI
reproduces Carnahan-Starling and the esFMT(1,0) gives something close to
Carnahan-Starling. For the homogeneous fluid at $\eta= 0.625$, this means that
Tarazona gives a free energy divided by temperature that is approximately
$0.84$ higher than the other two and this corresponds roughly to the
differences in the height of the first peaks and first minima of the free
energy curves while for $\eta= 0.655$ the difference is $1.1$ which again is
similar to the differences of the first peaks and first minima. It therefore
seems that the main difference between the various tensor models in the case
of the BCC system is that the Tarazona model goes to the less accurate and
somewhat higher PY free energy at $\alpha= 0$ and this continues to cause an
increase in the free energies at the low values of $\alpha$ at which the first
minimum occurs whereas it is forgotten at the high values of $\alpha$ where
the second minimum occurs and all three models give very similar results.

\section{Conclusion}

In this paper, the arguments leading to the most popular form of FMT, namely
the White Bear functionals, have been re-examined in light of the demand for
global stability of the functionals. The requirement that the functionals
reproduce the known exact results for a simple spherical cavity just large
enough to contain a single hard sphere (e.g. a primitive 0D cavity), for any
combination of two such cavities, for certain examples of three such cavities
and for a one-dimensional density distribution provide strong constraints on
the possible model functionals. Using the ansatz of Tarazona and Rosenfeld, as
well as demanding computationally practicality (e.g. limiting the complexity
of the three body kernel so that it is representable using scalar, vector and
second-order tensors measures only) leads to a two-parameter class of possible
functionals. If one further imposes that the first-order density correction to
the dcf of the homogeneous fluid is recovered, the parameters are fixed to
those originally suggested by Tarazona\cite{Tarazona} and that form the basis
of the White Bear functionals. The main observation of this work is that such
a choice cannot be proven to be globally stable (and, indeed, as noted above
there is strong evidence that it is not\cite{LutskoLam}) and so alternatives
were considered which do give global stability at the cost of less accuracy
at low density.

The proposed functional, labeled esFMT(1,0), is not only stable but seems
comparable to the WBI model in most numerical tests: it gives an equation of
state for the homogeneous system which is better than PY, if not quite as
accurate as Carnahan-Starling, it performs similarly for fluids near a wall
and the solid phase properties are all comparable to WBI. The main exception
to the last statement is in the vacancy concentration of the FCC solid where
the proposed model is in good agreement with simulation - better than WBII and
much better than WBI. This is, however, most likely a chance result and may
not hold any real significance in terms of the quality of the model. Like the
WB models, the explicitly stable FMT shows somewhat unphysical behavior for
the BCC phase and so this remains an overall weakness of the FMT approach.
Aside from the important issue of global stability, perhaps the biggest
conceptual virtue of the esFMT is that it achieves this positive comparison
with WB without introducing the heuristics of the WB models (which explicitly
build in the CS equation of state). One could also note that a practical
consequence of this is that the analytic form is somewhat simpler than that of
WB. Overall, the esFMT(1,0) model seems to be adequate for any current application.

In an ideal world, one would of course prefer to recover the low-density
limits as well but this would seem to only be possible if the model were
extended to higher tensorial-order - something which is not impossible, but
which will require further work. One would also like to have a more robust
criterion for the selection of the remaining parameters and here it may be
that something could be done using the ideas of Santos\cite{Santos} concerning
thermodynamic consistency of mixtures, which have yet to be fully incorporated
into FMT.

\begin{acknowledgments}
This work of JFL was supported by the European Space Agency (ESA) and the Belgian
Federal Science Policy Office (BELSPO) in the framework of the PRODEX
Programme, contract number ESA AO-2004-070.
\end{acknowledgments}

\appendix

\section{Exact results for 0D cavity}
\label{app0D}
To illustrate the calculations used to determine the free energy of
configurations of 0D cavities, consider first a cavity $\mathcal{S}$ in $D$
dimensions which is large enough to hold a single hard sphere but not large
enough to contain more than one without overlap. The grand partition function
is
\begin{equation}
\Xi\left[  \overrightarrow{\phi}\right]  =1+\Lambda^{-D}\int_{\mathcal{S}}%
\exp\left(  -\beta\left(  \phi\left(  \mathbf{q}_{1}\right)  -\mu\right)
\right)  d\mathbf{q}_{1}%
\end{equation}
where $\Lambda$ is the thermodynamic wavelength, $\phi\left(  \mathbf{r}%
\right)  $ is the external field (which is infinite outside the cavity but
arbitrary within) and $\mu$ is the chemical potential. There are only two
terms, corresponding to zero and one particles. The grand canonical free
energy is $\Omega\left[  \phi\right]  =-k_{B}T\ln\Xi\left[  \overrightarrow
{\phi}\right]  $ and the local density is given by the standard relation%
\begin{equation}
n\left(  \mathbf{r}\right)  =-\frac{\delta\Omega\left[  \phi\right]  }%
{\delta\phi\left(  \mathbf{r}\right)  }=\frac{\Lambda^{-D}\exp\left(
-\beta\left(  \phi\left(  \mathbf{r}\right)  -\mu\right)  \right)  }%
{1+\Lambda^{-D}\int_{\mathcal{S}}\exp\left(  -\beta\left(  \phi\left(
\mathbf{q}_{1}\right)  -\mu\right)  \right)  d\mathbf{q}_{1}} \label{a1}%
\end{equation}
This can be inverted to give the field in terms of the density
\begin{equation}
\exp\left(  -\beta\left(  \phi\left(  \mathbf{r}\right)  -\mu\right)  \right)
=\frac{\Lambda^{D}n\left(  \mathbf{r}\right)  }{1-\int_{V}n\left(
\mathbf{q}\right)  d\mathbf{q}}\equiv\frac{\Lambda^{D}n\left(  \mathbf{r}%
\right)  }{1-N\left[  n\right]  }%
\end{equation}
where the equivalence identifies the average number of particles in the cavity
which, by hypothesis, is $0\leq N\left[  n\right]  \leq1$. The cDFT free energy
functional is
\begin{align}
\beta F\left[  n\right]   &  =\left(  \beta\Omega\left[  \phi\right]
-\beta\int_{V}n\left(  \mathbf{r}\right)  \left(  \phi\left(  \mathbf{r}%
\right)  -\mu\right)  d\mathbf{r}\right)  _{\phi\left[  n\right]  }%
\label{0D}\\
&  =\beta F_{id}\left[  n\right]  +\left(  1\mathbf{-}N\left[  n\right]
\right)  \ln\left(  1-N\left[  n\right]  \right)  +N\left[  n\right] \nonumber
\end{align}
For a spherical cavity of radius $R+\varepsilon$ and with zero field inside
the cavity, and writing the volume of a D-dimensional spherical cavity of radius $R$ as  $V_{D}(R)$, Eq.(\ref{a1}) gives
\begin{equation}
n_{\varepsilon}\left(  \mathbf{r}\right)  =\frac{\Lambda^{-D}\Theta\left(
\varepsilon-r\right)  }{1+\Lambda^{-D}V_{D}\left(  \varepsilon\right)
\exp\left(  -\beta\mu\right)  }=N\left[  n\right]  \frac{\Theta\left(
\varepsilon-r\right)  }{V_{D}\left(  \varepsilon\right)  }%
\end{equation}
which shows that $\lim_{\varepsilon\rightarrow0}n_{\varepsilon}\left(
\mathbf{r}\right)  =N\delta\left(  \mathbf{r}\right)  $. The calculation
generalizes trivially for any number of cavities that do not intersect as well
as the case that they have a non-zero mutual intersection. For other cases,
the results become more complex. For example, for three spherical cavities
having the property that $\mathcal{S}_{1}\cap\mathcal{S}_{2}\neq\emptyset
\neq\mathcal{S}_{2}\cap\mathcal{S}_{3}$ but $\mathcal{S}_{1}\cap
\mathcal{S}_{3}=\emptyset$, one finds
\begin{align}
n\left(  \mathbf{r}\right)   &  =\left(  \frac{x+x^{2}}{1+3x+x^{2}}\right)
V_{D}^{-1}\left(  \varepsilon\right)  \left(  \Theta\left(  \varepsilon
-\left\vert \mathbf{r-S}_{1}\right\vert \right)  +\Theta\left(  \varepsilon
-\left\vert \mathbf{r-S}_{3}\right\vert \right)  \right)  +\frac{x}%
{1+3x+x^{2}}V_{D}^{-1}\left(  \varepsilon\right)  \Theta\left(  \varepsilon
-\left\vert \mathbf{r-S}_{2}\right\vert \right) \\
x  &  =\frac{3\left\langle N;\left[  n\right]  \right\rangle -3+\sqrt
{5\left\langle N;\left[  n\right]  \right\rangle ^{2}-10\left\langle N;\left[
n\right]  \right\rangle +9}}{4-2\left\langle N;\left[  n\right]  \right\rangle
}\nonumber
\end{align}
and the Helmholtz free energy functional becomes correspondingly nontrivial.

\section{Evaluating FMT for zero-dimensional cavities}
\label{appB}

\subsection{The packing fraction}

Consider a collection of $m$ 0D-cavities with centers $\mathbf{s}_{i}$. Note that in the following, everything is written in terms of the limit that the density becomes a sum of delta functions but the same manipulations can be made, more mathematically securely, with cavities slightly larger than a sphere for which the density is not singular (as described in the previous Appendix) with the same results in the singular limit.

The local density is therefore
\begin{equation}
n\left(  \mathbf{r}\right)  =\sum_{i=1}^{m}N_{i}\delta\left(  \mathbf{r-s}%
_{i}\right)  ,
\end{equation}
where the restrictions on the coefficients depend on the geometry: if the
cavities are all disjoint, the $0\leq N_{i}\leq1$, if there is a non-empty
mutual intersection, then $0\leq\sum_{i=1}^{m}N_{i}\leq N$, etc. . The local
packing fraction is
\begin{equation}
\eta\left(  \mathbf{r}\right)  =\sum_{i=1}^{m}N_{i}\Theta\left(  R-\left\vert
\mathbf{r-s}_{i}\right\vert \right)
\end{equation}
and one verifies that for any function $\psi\left(  x\right)  $,
\begin{equation}
\frac{\partial}{\partial R}\psi\left(  \eta\left(  \mathbf{r}\right)  \right)
=\psi^{\prime}\left(  \eta\left(  \mathbf{r}\right)  \right)  \sum_{i=1}%
^{m}N_{i}\delta\left(  R-\left\vert \mathbf{r-s}_{i}\right\vert \right)  .
\end{equation}
Furthermore, if one writes
\begin{equation}
\eta\left(  \mathbf{r}\right)  =\lim_{R_{1}...R_{m}\rightarrow R}\sum
_{i=1}^{m}N_{i}\Theta\left(  R_{i}-\left\vert \mathbf{r-s}_{i}\right\vert
\right)
\end{equation}
then
\begin{align}
\frac{\partial}{\partial R_{i}}\psi\left(  \eta\left(  \mathbf{r}\right)
\right)   &  =\psi^{\prime}\left(  \eta\left(  \mathbf{r}\right)  \right)
N_{i}\delta\left(  R_{i}-\left\vert \mathbf{r-s}_{i}\right\vert \right) \\
\frac{\partial^{2}}{\partial R_{i}\partial R_{j}}\psi\left(  \eta\left(
\mathbf{r}\right)  \right)   &  =\psi^{\prime\prime}\left(  \eta\left(
\mathbf{r}\right)  \right)  N_{i}N_{j}\delta\left(  R_{i}-\left\vert
\mathbf{r-s}_{i}\right\vert \right)  \delta\left(  R_{j}-\left\vert
\mathbf{r-s}_{j}\right\vert \right)  +\delta_{ij}\psi^{\prime}\left(
\eta\left(  \mathbf{r}\right)  \right)  N_{i}\delta^{\prime}\left(
R_{i}-\left\vert \mathbf{r-s}_{i}\right\vert \right) \nonumber
\end{align}
and so forth.

\subsection{A lemma}

We will also need to evaluate terms of the form
\begin{equation}
\int\psi\left(  \eta\left(  \mathbf{r};\left[  \rho\right]  \right)  \right)
d\mathbf{r}%
\end{equation}
which in general will depend on the geometry. To do so, let the volume of the
sphere centered at $\mathbf{s}_{i}$ and having radius $R_{i}$ be $V_{i}%
=V_{D}\left(  R_{i}\right)  $, let the volume of the intersection between
spheres $i$ and $j$ be $V_{ij}$, the mutual intersection of $i,j$ and $k$ be
$V_{ijk}$, etc. Finally, let $\widetilde{V}_{i}$ be the sub-volume of cavity
$i$ which is not simultaneous part of any other cavity; $\widetilde{V}_{ij}$
the sub-volume of the intersection of cavities $i$ and $j$ which is not in the
intersection with any third cavity, etc. For three cavities,
\begin{align}
\widetilde{V}_{123}  &  =V_{123}\\
\widetilde{V}_{12}  &  =V_{12}-\widetilde{V}_{123}=V_{12}-V_{123}\nonumber\\
\widetilde{V}_{1}  &  =V_{1}-\widetilde{V}_{12}-\widetilde{V}_{13}%
-\widetilde{V}_{123}=V_{1}-V_{12}-V_{13}+V_{123}\nonumber
\end{align}

Then, for $m$ cavities, one has that
\begin{align}
\int\psi\left(  \eta\left(  \mathbf{r};\left[  \rho\right]  \right)  \right)
d\mathbf{r}  &  = \sum_{i=1}^{m}\widetilde{V}_{i}\psi\left(  N_{i}\right) \\
&  +\sum_{i<j=1}^{m}\widetilde{V}_{ij}\psi\left(  N_{i}+N_{j}\right)
\nonumber\\
&  +\sum_{i<j<k=1}^{m}\widetilde{V}_{ijk}\psi\left(  N_{i}+N_{j}+N_{k}\right)
\nonumber\\
&  +...\nonumber
\end{align}
and specializing to $m=3$, this can be written as
\begin{align}
\int\psi\left(  \eta\left(  \mathbf{r};\left[  \rho\right]  \right)  \right)
d\mathbf{r}  &  =V_{123}\left(  \psi\left(  N_{1}+N_{2}+N_{3}\right)
-\sum_{i<j=1}^{3}\psi\left(  N_{i}+N_{j}\right)  +\sum_{i=1}^{3}\psi\left(
N_{i}\right)  \right) \label{psi}\\
&  +\sum_{i<j=1}^{3}V_{ij}\left(  \psi\left(  N_{i}+N_{j}\right)  -\psi\left(
N_{i}\right)  -\psi\left(  N_{j}\right)  \right) \nonumber\\
&  +\sum_{i=1}^{3}V_{i}\psi\left(  N_{i}\right) \nonumber
\end{align}
Note that the case $m=2$ follows by setting $V_{123}=V_{13}=V_{23}=0$ and that
of $m=1$ is just $V_{1}\psi\left(  N_{1}\right)  $.

\subsection{Evaluation of $F_{ex}[n]$}

Now, consider the first FMT term,%
\begin{align}
F_{\text{ex}}^{\left(  1\right)  }\left[  \rho\right]   &  =\int
d\mathbf{r}\;\psi_{1}\left(  \eta\left(  \mathbf{r};\left[  \rho\right]
\right)  \right)  \int d\mathbf{r}_{1}\;\rho\left(  \mathbf{r-r}_{1}\right)
\delta\left(  R-r_{1}\right)  K_{1}\left(  \mathbf{r}_{1}\right) \\
&  =\int d\mathbf{r}\;\psi_{1}\left(  \eta\left(  \mathbf{r};\left[
\rho\right]  \right)  \right)  \sum_{i=1}^{m}N_{i}\delta\left(  R-\left\vert
\mathbf{r-s}_{i}\right\vert \right)  K_{1}\left(  \widehat{\mathbf{r}}%
_{1}\right) \nonumber\\
&  =\overline{K}_{1}\frac{\partial}{\partial R}\int d\mathbf{r}\;\psi
_{0}\left(  \eta\left(  \mathbf{r};\left[  \rho\right]  \right)  \right)
\nonumber
\end{align}
for some function $\psi_{0}\left(  x\right)  $ satisfying $\psi_{1}\left(
x\right)  =\psi_{0}^{\prime}\left(  x\right)  $ and where it is noted that
since $K_{1}$ is a scalar, it must be a constant. Using Eq.(\ref{psi}) for the
case of $m=3$, this gives%
\begin{align}
F_{\text{ex}}^{\left(  1\right)  }\left[  \rho\right]   &  =\overline{K}%
_{1}\frac{\partial}{\partial R}V_{123}\left(  \psi\left(  N_{1}+N_{2}%
+N_{3}\right)  -\sum_{i<j=1}^{3}\psi\left(  N_{i}+N_{j}\right)  +\sum
_{i=1}^{3}\psi\left(  N_{i}\right)  \right) \\
&  +\overline{K}_{1}\sum_{i<j=1}^{3}\frac{\partial}{\partial R}V_{ij}\left(
\psi\left(  N_{i}+N_{j}\right)  -\psi\left(  N_{i}\right)  -\psi\left(
N_{j}\right)  \right) \nonumber\\
&  +\overline{K}_{1}\frac{\partial}{\partial R}V\left(  R\right)  \sum
_{i=1}^{3}\psi\left(  N_{i}\right) \nonumber
\end{align}

For the two-body contribution, one requires
\begin{align}
F_{\text{ex}}^{\left(  2\right)  }\left[  \rho\right]   &  =\int
d\mathbf{r}\;\psi_{2}\left(  \eta\left(  \mathbf{r};\left[  \rho\right]
\right)  \right)  \int d\mathbf{r}_{1}d\mathbf{r}_{2}\;\rho\left(
\mathbf{r-r}_{1}\right)  \rho\left(  \mathbf{r-r}_{2}\right)  \delta\left(
R-r_{1}\right)  \delta\left(  R-r_{2}\right)  K_{2}\left(  \mathbf{r}%
_{1},\mathbf{r}_{2}\right) \\
&  =\sum_{i,j=1}^{m}N_{i}N_{j}\int d\mathbf{r}\;\psi_{2}\left(  \eta\left(
\mathbf{r};\left[  \rho\right]  \right)  \right)  \delta\left(  R-\left\vert
\mathbf{r-s}_{i}\right\vert \right)  \delta\left(  R-\left\vert \mathbf{r-s}%
_{j}\right\vert \right)  K_{2}\left(  \mathbf{r-s}_{i},\mathbf{r-s}_{j}\right)
\nonumber
\end{align}
and noting that $K_{2}\left(  \mathbf{r}_{1},\mathbf{r}_{2}\right)  $ can only
be a function of $\mathbf{r}_{1}\cdot\mathbf{r}_{2}$ and that in general for
any function $f\left(  x\right)  $
\begin{equation}
\delta\left(  R-r_{1}\right)  \delta\left(  R-r_{2}\right)  f\left(
\mathbf{r}_{1}\cdot\mathbf{r}_{2}\right)  =\delta\left(  R-r_{1}\right)
\delta\left(  R-r_{2}\right)  f\left(  \frac{r_{1}^{2}+r_{2}^{2}-\left(
\mathbf{r}_{1}-\mathbf{r}_{2}\right)  ^{2}}{2}\right)  =\delta\left(
R-r_{1}\right)  \delta\left(  R-r_{2}\right)  f\left(  R^{2}\left(
1-\frac{\left(  \mathbf{r}_{1}-\mathbf{r}_{2}\right)  ^{2}}{2R^{2}}\right)
\right)
\end{equation}
so
\begin{equation}
F_{\text{ex}}^{\left(  2\right)  }\left[  \rho\right]  =\sum_{i,j=1}%
^{m}\overline{K}_{2}\left(  1-\frac{s_{ij}^{2}}{2R^{2}}\right)  \int
d\mathbf{r}\;\psi_{2}\left(  \eta\left(  \mathbf{r};\left[  \rho\right]
\right)  \right)  N_{i}N_{j}\delta\left(  R-\left\vert \mathbf{r-s}%
_{i}\right\vert \right)  \delta\left(  R-\left\vert \mathbf{r-s}%
_{j}\right\vert \right)
\end{equation}
for some function $\overline{K}_{2}\left(  x\right)  $. Assuming that
$\overline{K}_{2}\left(  1\right)  =0$, as discussed in the main text, this
gives
\begin{equation}
F_{\text{ex}}^{\left(  2\right)  }\left[  \rho\right]  =\sum_{i,j=1}%
^{m}\overline{K}_{2}\left(  1-\frac{s_{ij}^{2}}{2R^{2}}\right)  \lim
_{R_{i},R_{j}\rightarrow R}\frac{\partial^{2}}{\partial R_{i}\partial R_{j}%
}\int\psi_{0}\left(  \eta\left(  \mathbf{r};\left[  \rho\right]  \right)
\right)  \;d\mathbf{r}%
\end{equation}
For $m=2$, this becomes
\begin{equation}
  \label{B16}
F_{\text{ex}}^{\left(  2\right)  }\left[  \rho\right]  =\left(  \psi\left(
N_{1}+N_{2}\right)  -\psi\left(  N_{1}\right)  -\psi\left(  N_{2}\right)
\right)  2\overline{K}_{2}\left(  1-\frac{s_{ij}^{2}}{2R^{2}}\right)
\lim_{R_{1},R_{2}\rightarrow R}\frac{\partial^{2}}{\partial R_{1}\partial
R_{2}}V_{12}%
\end{equation}
The evaluation of $F_{\text{ex}}^{\left(  3\right)  }\left[  \rho\right]  $
proceeds similarly.

\section{Reproducing the 1D functional\label{1D}}

In this Appendix, it is shown that as long as the three-body (and higher
order) kernels satisfy the stability condition (i.e. they vanish whenever two
arguments are the same) they can give no contribution to 1D dimensional
crossover and, so, preserve the Percus functional. To see this, consider the
n-body term%
\begin{equation}
F_{\text{ex}}^{\left(  n\right)  }\left[  \rho\right]  =\int d\mathbf{r}%
\;\psi_{n}\left(  \eta\left(  \mathbf{r};\left[  \rho\right]  \right)
\right)  \int d\mathbf{r}_{1}...d\mathbf{r}_{n}\;\rho\left(  \mathbf{r-r}%
_{1}\right)  ...\rho\left(  \mathbf{r-r}_{n}\right)  \delta\left(
R-r_{1}\right)  ...\delta\left(  R-r_{n}\right)  K_{n}\left(  \widehat
{\mathbf{r}}_{1},...,\widehat{\mathbf{r}}_{n}\right)
\end{equation}
and the restriction of the density to one dimension%
\begin{equation}
\rho\left(  \mathbf{r}\right)  =\delta\left(  x\right)  \delta\left(
y\right)  \rho\left(  z\right)
\end{equation}
so that%
\begin{align}
\rho\left(  \mathbf{r-r}_{i}\right)  \delta\left(  R-r_{i}\right)    &
=\delta\left(  x-x_{i}\right)  \delta\left(  y-y_{i}\right)  \rho\left(
z-z_{i}\right)  \delta\left(  R-\sqrt{x_{i}^{2}+y_{i}^{2}+z_{i}^{2}}\right)
\\
& =\delta\left(  x-x_{i}\right)  \delta\left(  y-y_{i}\right)  \rho\left(
z-z_{i}\right)  \delta\left(  R-\sqrt{x^{2}+y^{2}+z_{i}^{2}}\right)
\nonumber\\
& =\delta\left(  x-x_{i}\right)  \delta\left(  y-y_{i}\right)  \rho\left(
z-z_{i}\right)  \left[  \frac{R}{\left\vert z_{i}\right\vert }\delta\left(
z_{i}-\sqrt{R^{2}-x^{2}-y^{2}}\right)  +\frac{R}{\left\vert z_{i}\right\vert
}\delta\left(  z_{i}+\sqrt{R^{2}-x^{2}-y^{2}}\right)  \right]  \Theta\left(
R^{2}-x^{2}-y^{2}\right)  \nonumber\\
& =\left[  \frac{R}{\sqrt{R^{2}-x^{2}-y^{2}}}\rho\left(  z-\sqrt{R^{2}%
-x^{2}-y^{2}}\right)  \Theta\left(  R^{2}-x^{2}-y^{2}\right)  \right]
\nonumber\\
& \times\delta\left(  x-x_{i}\right)  \delta\left(  y-y_{i}\right)  \left[
\delta\left(  z_{i}-\sqrt{R^{2}-x^{2}-y^{2}}\right)  +\delta\left(
z_{i}+\sqrt{R^{2}-x^{2}-y^{2}}\right)  \right]  \nonumber
\end{align}
This means that the kernel is evaluated with
\begin{equation}
\widehat{\mathbf{r}}_{i}=\frac{\mathbf{r}_{i}}{R}=\frac{x\widehat{\mathbf{x}%
}+y\widehat{\mathbf{y}}\pm\sqrt{R^{2}-x^{2}-y^{2}}\widehat{\mathbf{z}}}{R}%
\end{equation}
so that for $n\geq3$  it is always the case that at least two are the same and
so the kernel vanishes. Thus, these higher order terms cannot affect the
crossover to 1D. 

\bibliography{stable}

\end{document}